\documentclass{aa}

\usepackage{graphicx}
\usepackage[varg]{txfonts}
\usepackage{natbib}
\usepackage[pdftex]{hyperref}
\usepackage{array}
\def\ImUnit{\mathbf{i}}

\def\sun{\odot}

\def\expo#1{\mathbf{e}^{#1}}

\graphicspath{{./figures/}}

\newcommand\figI{
  \begin{figure}
    \centering
    \includegraphics[width=\linewidth]{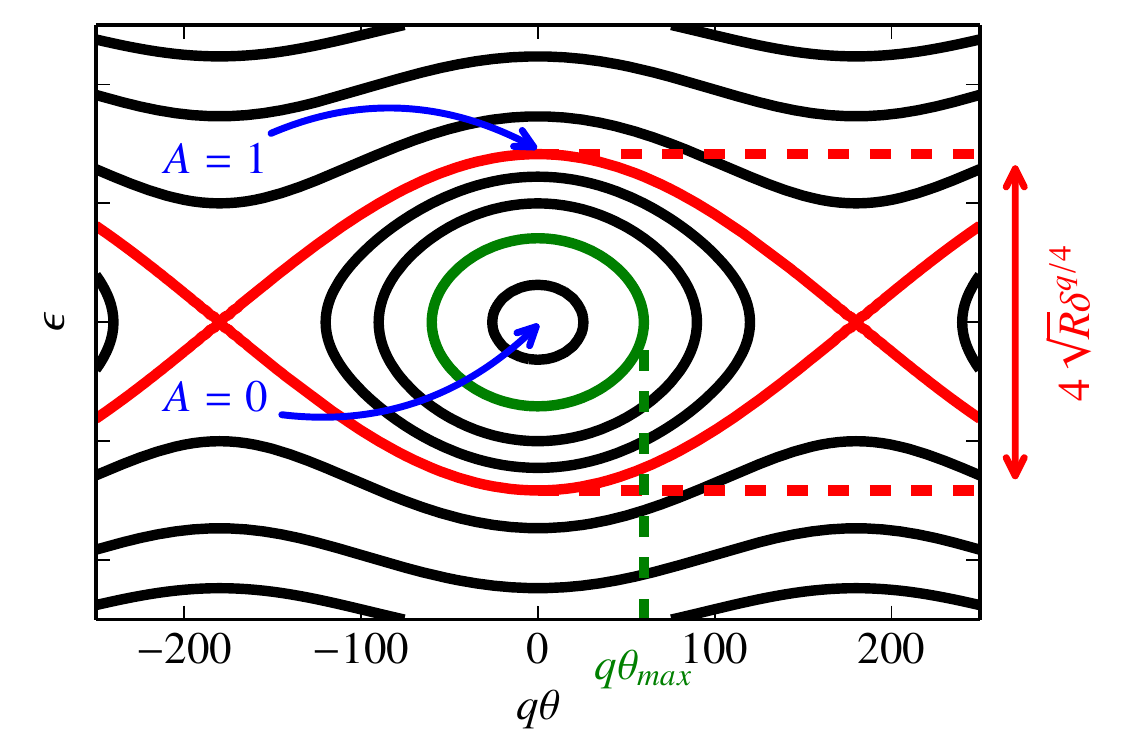}
    \caption{Phase space of a resonance of order $q$
      in the simplified pendulum-like approximation
      (Hamiltonian~(\ref{eq:Hqeps})).
      $\theta$ is the unique resonant angle and
      $\epsilon$ its conjugated action.
      The separatrix is highlighted in red.
      The amplitude $A$ (defined with $\theta_{max}$ see Eq.~(\ref{eq:Ampli}))
      is 0 at the center
      of the resonance (elliptical fixed point) and 1 at the separatrix.
    }
    \label{fig:I}
  \end{figure}
}

\newcommand\figII{
  \begin{figure*}
    \centering
    \includegraphics[width=18cm]{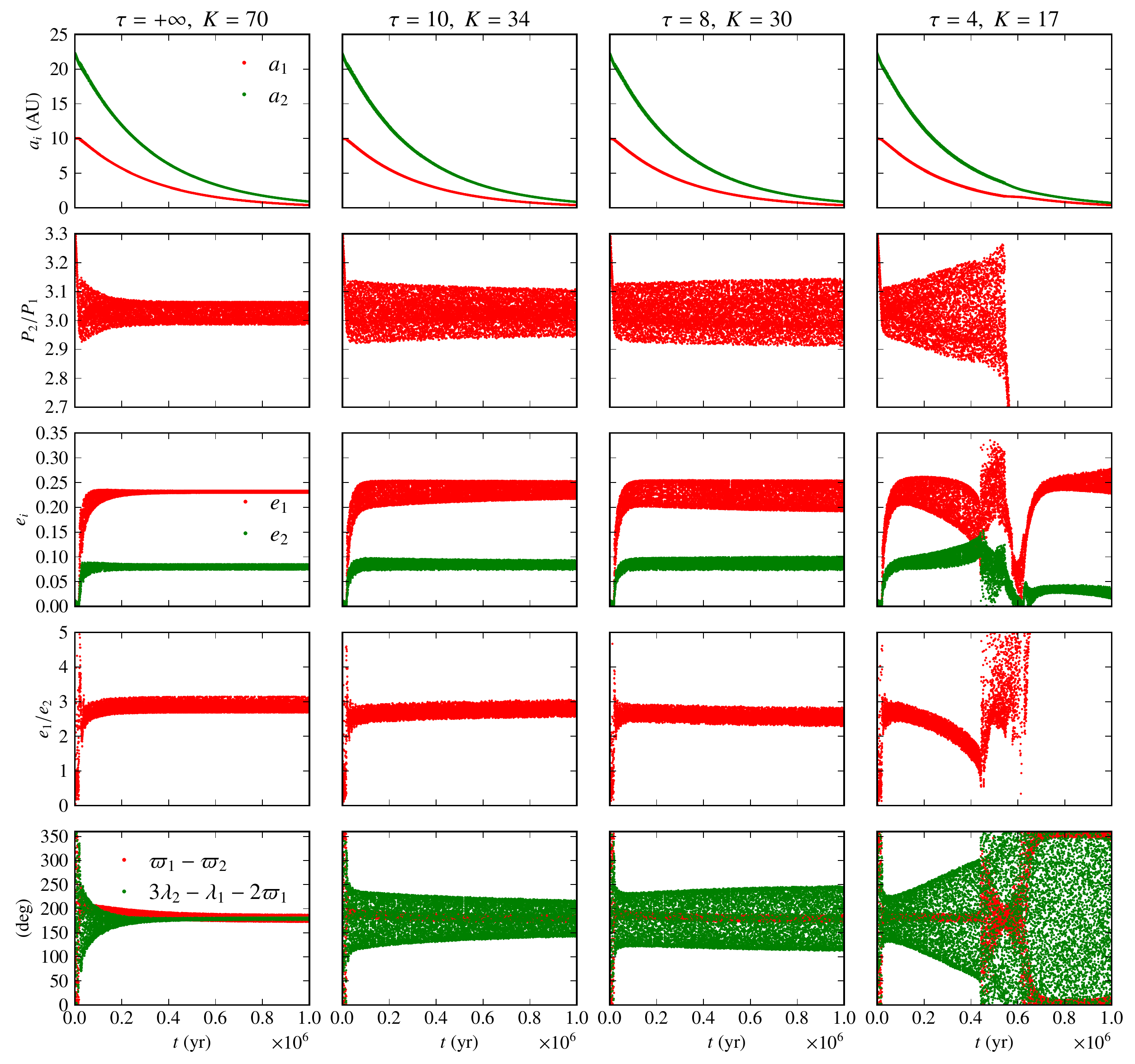}
    \caption{Semi-major axes, period ratio, eccentricities, eccentricity ratio,
      and angles evolution for simulations of \object{HD~60532}b, c
      with different dissipation timescale ratios $\tau=T_{e,1}/T_{e,2}=T_{m,1}/T_{m,2}$.
      The ratio $K=T_{m,i}/T_{e,i}$ is set according to Eq.~(\ref{eq:constK}) to
      reproduce the observed equilibrium eccentricities.
      We used $\tau = +\infty$, 10, 8, 4 with K = 70, 34, 30, 17 respectively
      for the four shown simulations (four columns).
      The amplitude of libration decreases for the first two simulations
      ($\tau=+\infty$, 10) and increases for the last two ($\tau=8$, 4).
      The value given by our analytical criterion for the transition
      between decreasing and increasing amplitude is $\tau\sim 9$.
    }
    \label{fig:II}
  \end{figure*}
}

\newcommand\figIII{
  \begin{figure*}
    \centering
    \includegraphics[width=18cm]{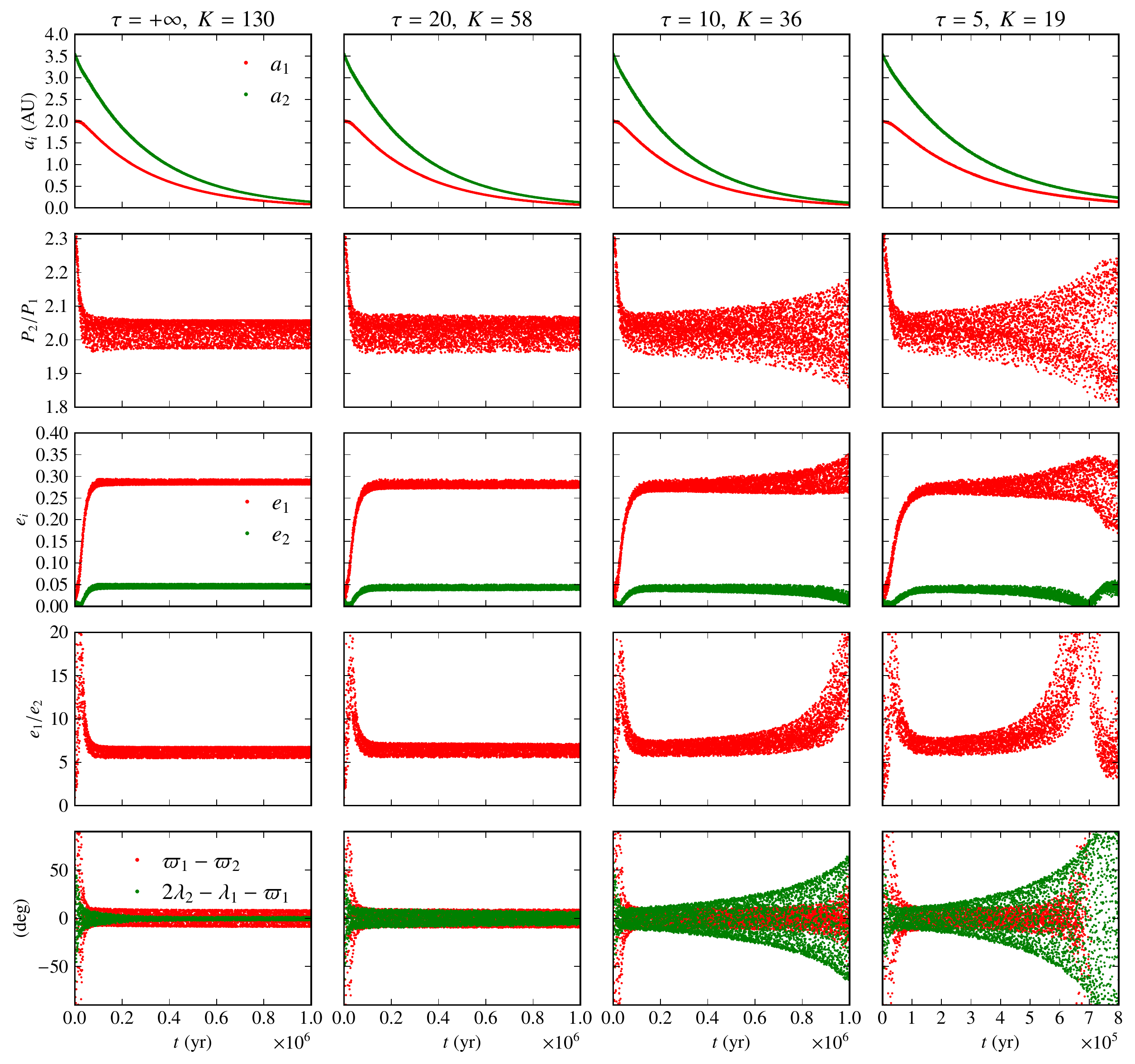}
    \caption{Same as Fig.~\ref{fig:II} but for \object{GJ~876}b, c.
      We used $\tau = +\infty$, 20, 10, 5 with K = 130, 58, 36, 19 respectively
      for the four shown simulations (four columns).
      Note that the last simulation ($\tau=5$, $K=19$) ended before $10^6$~yr (around $8\times 10^5$~yr) because of orbital
      instability when the system escaped from resonance.
      The amplitude of libration decreases for the first two simulations
      ($\tau=+\infty$, 20) and increases for the last two ($\tau=10$, 5).
      The value given by our analytical criterion for the transition
      between decreasing and increasing amplitude is $\tau\sim 42$.
    }
    \label{fig:III}
  \end{figure*}
}

\newcommand\figIV{
  \begin{figure*}
    \centering
    \includegraphics[width=18cm]{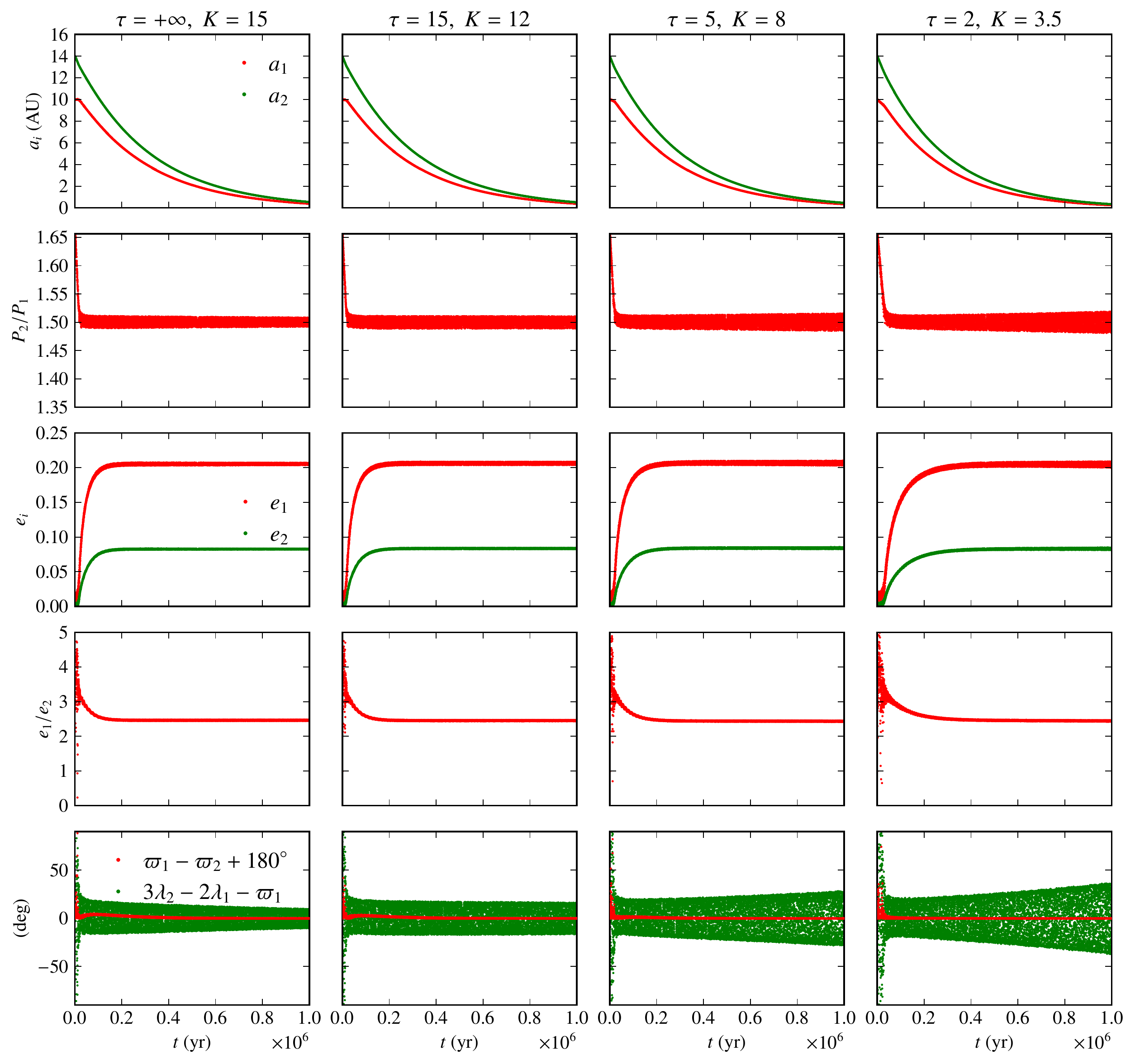}
    \caption{Same as Fig.~\ref{fig:II} but for \object{HD~45364}b, c.
      We used $\tau = +\infty$, 15, 5, 2 with K = 15, 12, 8, 3.5 respectively
      for the four shown simulations (four columns).
      The amplitude of libration decreases for the first two simulations
      ($\tau=+\infty$, 15) and increases for the last two ($\tau=5$, 2).
      The value given by our analytical criterion for the transition
      between decreasing and increasing amplitude is $\tau\sim 6.3$.
    }
    \label{fig:IV}
  \end{figure*}
}

\newcommand\tabI{
  \begin{table}
    \begin{center}
      \caption{Orbital parameters of \object{HD~60532}b,c used in this study
        \citep[taken from][]{laskar_planetary_2009}.
        The stellar mass is $1.44 M_\sun$.}
      \begin{tabular}{cc|cc}
        \hline
        Parameter & [unit] & b & c\\
        \hline
        $m$ & [$M_J$] &
        $3.1548$ & $7.4634$ \\
        $P$ & [day] &
        $201.83$ & $607.06$ \\
        $a$ & [AU] &
        $0.7606$ & $1.5854$ \\
	$e$ & &
        $0.278$ & $0.038$ \\
        \hline
      \end{tabular}
      \label{tab:I}
    \end{center}
  \end{table}
}

\newcommand\tabII{
  \begin{table}
    \begin{center}
      \caption{Orbital parameters of \object{GJ~876}b,c used in this study
        \citep[taken from][]{correia_harps_2010}.
        The stellar mass is $0.334 M_\sun$.}
      \begin{tabular}{cc|cc}
        \hline
        Parameter & [unit] & c & b\\
        \hline
        $m$ & [$M_J$] &
        $0.86$ & $2.64$ \\
        $P$ & [day] &
        $30.259$ & $61.065$ \\
        $a$ & [AU] &
        $0.132$ & $0.211$ \\
	$e$ & &
        $0.265$ & $0.031$ \\
        \hline
      \end{tabular}
      \label{tab:II}
    \end{center}
  \end{table}
}

\newcommand\tabIII{
  \begin{table}
    \begin{center}
      \caption{Orbital parameters of \object{HD45364}b,c used in this study
        \citep[taken from][]{correia_harps_2009}.
        The stellar mass is $0.82 M_\sun$.}
      \begin{tabular}{cc|cc}
        \hline
        Parameter & [unit] & c & b\\
        \hline
        $m$ & [$M_J$] &
        $0.1872$ & $0.6579$ \\
        $P$ & [day] &
        $226.93$ & $342.85$ \\
        $a$ & [AU] &
        $0.6813$ & $0.8972$ \\
	$e$ & &
        $0.1684$ & $0.0974$ \\
        \hline
      \end{tabular}
      \label{tab:III}
    \end{center}
  \end{table}
}

\begin{document}

\title{Stability of resonant configurations during the migration of planets and constraints on disk-planet interactions}
\titlerunning{Stability of resonances during migration}
\author{J.-B. Delisle\inst{1,2}
  \and A. C. M. Correia\inst{2,3}
  \and J. Laskar\inst{2}}

\institute{Observatoire de l'Université de Genève, 51 chemin des Maillettes, 1290, Sauverny, Switzerland\\
  \email{jean-baptiste.delisle@unige.ch}
  \and ASD, IMCCE-CNRS UMR8028, Observatoire de Paris, UPMC,
  77 Av. Denfert-Rochereau, 75014~Paris, France
  \and CIDMA, Departamento de F\'isica, Universidade de Aveiro, Campus de Santiago,
  3810-193 Aveiro, Portugal
}

\date{\today}

\abstract{
We study the stability of mean-motion resonances (MMR) between two planets
during their migration in a protoplanetary disk.
We use an analytical model of resonances, and describe the effect of the disk
by a migration timescale ($T_{m,i}$) and an eccentricity damping timescale ($T_{e,i}$)
for each planet ($i=1,2$ respectively for the inner and outer planet).
We show that the resonant configuration is stable if
$T_{e,1}/T_{e,2} > (e_1/e_2)^2$.
This general result can be
used to put constraints
on specific models of disk-planet interactions.
For instance, using classical prescriptions for type I migration, we show that
when the angular momentum deficit (AMD) of the inner
orbit is larger than the outer's orbit AMD, resonant systems must have
a locally inverted disk density profile
to stay locked in resonance during the migration.
This inversion is very untypical of type I migration and our
criterion can thus provide an evidence against classical type I migration.
That is indeed the case for the Jupiter-mass resonant systems
\object{HD~60532}b, c (3:1 MMR),
\object{GJ~876}b, c (2:1 MMR), and \object{HD~45364}b, c (3:2 MMR).
This result may be an evidence for type II migration (gap opening planets),
which is compatible with the large masses of these planets.
}


\maketitle


\section{Introduction}
\label{sec:introduction}

In \citet{delisle_resonance_2014}, we showed that tidal dissipation raised by the star
on two resonant planets can produce three kinds of distinct evolutions
depending on the relative strength of the dissipation in both planets.
The three different outcomes of this tidal process are systems that stay in resonance,
systems that leave the resonance with an increasing period ratio ($P_{out}/P_{in}$)
and systems that leave the resonance with a decreasing period ratio.
For known near resonant systems, the comparison of the period ratio of the planets
with respect to the nominal resonant value
helps to put constraints on the tidal dissipation undergone by each planet
and thus on the nature of the planets \citep[see][]{delisle_resonance_2014}.
In this article, we generalize our reasoning to other forms of dissipation,
in particular to disk-planet interactions.

Disk-planet interactions can induce migration of the planets \citep[e.g.][]{goldreich_excitation_1979}.
In the case of convergent migration (i.e. decreasing period ratio),
the planets can be locked in resonance \citep[e.g.][]{weidenschilling_orbital_1985}.
Two planets that are locked in resonance have their eccentricities excited on the migration timescale
\citep[e.g.][]{weidenschilling_orbital_1985}.
However, disk-planet interactions also induce exponential eccentricity damping.
Depending on the respective timescales of the migration and eccentricity damping,
the system can reach a stationary state in which eccentricities stay constant \citep[][]{lee_dynamics_2002}.
The semi-major axes continue to evolve but the semi-major axis ratio (or period ratio)
stays locked at the resonant value.
Recently, \citet{goldreich_overstable_2014} showed that this equilibrium is unstable
in the case of the circular restricted three body problem where the inner planet has negligible mass.
This means that after the resonance locking, the eccentricity of the inner planet reach an equilibrium value
but then undergoes larger and larger oscillations around this equilibrium value
until the system reaches the resonance separatrix and leaves the resonance.
Then, the period ratio is no more locked at the resonant value
and the convergent migration continues (decreasing period ratio) until the system reaches another resonance.
The timescale of the resonance escape is given by the eccentricity damping timescale and
is thus short compared to the migration timescale \citep[see][]{goldreich_overstable_2014}.
Therefore, \citet{goldreich_overstable_2014} concluded that when the disk disappears and the migration stops,
only a few systems should be observed in resonance.
However, this conclusion is mainly based on a particular case in which the mass of the inner planet
is much smaller than the mass of the outer planet whose eccentricity is negligible and the migration
and damping forces are only undergone by the inner planet.
As shown in \citet{delisle_resonance_2014}, the evolution of a resonant system under dissipation
highly depends on which planet is affected by the dissipation.
In this paper we study a more general case in which both planets have masses,
eccentricities, and undergo dissipative forces.

In Sect.~\ref{sec:cons} we introduce the notations and the model of the resonant motion
in the conservative case that we developed in
\citet{delisle_resonance_2014}.
In Sect.~\ref{sec:dissip} we study the dissipative evolution of resonant planets
in a very general framework (Sect.~\ref{sec:general}), and we apply this modeling
to disk-planet interactions (Sect.~\ref{sec:disk}).
In Sect.~\ref{sec:applications} we show how our model can be used to put constraints on disks properties
for observed resonant systems.
In Sect.~\ref{sec:appli} we apply these analytical constraints to selected examples, and
compare them with numerical simulations.
We specifically study \object{HD~60532}b, c (3:1 resonance, Sect.~\ref{sec:hd60532})
\object{GJ~876}b, c (2:1 resonance, Sect.~\ref{sec:gj876}),
and \object{HD~45364}b, c (3:2 resonance, Sect.~\ref{sec:hd45364}).

\section{Resonant motion in the conservative case}
\label{sec:cons}

In the following, we refer to the star as body 0, to the inner planet as body 1, and to the outer planet as body 2.
We note $m_i$ the masses of the three bodies, and introduce
$\mu_i = \mathcal{G} (m_0 + m_i)$
and $\beta_i = m_0 m_i/(m_0 + m_i)$,
where $\mathcal{G}$ is the gravitational constant.
We only consider the planar case in this study.

In \citet{delisle_resonance_2014} we constructed a simplified,
and integrable model of the resonant motion in the conservative and planar case.
The main simplification of this model is to assume that the eccentricity ratio
($e_1/e_2$) stays close to the forced eccentricities ratio ($e_{1,ell}/e_{2,ell}$).
These forced eccentricities correspond to the eccentricities at the elliptical fixed point
at the resonance libration center.
With this assumption, and assuming moderate eccentricities
\footnote{The pendulum approximation of resonances
  is obtained using an analytical expansion in power series of eccentricities
  and is thus not valid at high eccentricities.
  Moreover, when eccentricities are vanishing,
  the phase space bifurcate and a better approximation is given
  by the second fundamental model of resononaces
  \citep[see][]{henrard_second_1983,delisle_dissipation_2012}.},
the Hamiltonian of the system can be simplified \citep[see][]{delisle_resonance_2014}
to the following simple pendulum Hamiltonian
\begin{equation}
  \label{eq:Hqeps}
  \mathcal{H} = -\epsilon^2 + 2 R \delta^{q/2} \cos(q\theta),
\end{equation}
where $q$ is the degree of the resonance ($q = k_2-k_1$ for a $k_2$:$k_1$ resonance),
$\epsilon$ is the action coordinate and provides a measure of the distance to the exact commensurability.
$\theta$ is the unique resonant angle in this simplified model.
It is a combination of both usual resonant angles
\citep[$\sigma_i = \frac{k_2}{q}\lambda_2-\frac{k_1}{q}\lambda_1-\varpi_i$, see Appendix~\ref{sec:simplif},
Eqs.~(\ref{eq:uDtheta}) and (\ref{eq:u}) and][]{delisle_resonance_2014}.
$R$ is a constant that depends on the masses of the bodies and on the considered resonance
\citep[see][]{delisle_resonance_2014}.
$\delta$ is a constant of motion (parameter of the model).
We have
\begin{eqnarray}
  \label{eq:epsilon}
  \epsilon &=& \Lambda_1-\Lambda_{1,0} + \Lambda_2-\Lambda_{2,0},\\
  \label{eq:delta}
  \delta &=& \Lambda_{1,0}-G_1 + \Lambda_{2,0}-G_2,
\end{eqnarray}
where
$\Lambda_i$ is the renormalized circular angular momentum of planet $i$, $G_i$ its
renormalized angular momentum \citep[see Appendix~\ref{sec:renorm} and][]{delisle_resonance_2014}.
The subscript 0 denotes the values at the exact commensurability.
The quantities $\Lambda_i$ only depend on the semi-major axis ratio $\alpha = a_1/a_2$
\begin{eqnarray}
  \label{eq:La1alpha}
  \Lambda_1(\alpha) &=& \frac{1}{(k_2/k_1) + (\beta_2/\beta_1)\sqrt{\mu_2/(\mu_1\alpha})}\\
  &\approx& \frac{1}{(k_2/k_1) + (m_2/m_1)/\sqrt{\alpha}}, \nonumber\\
  \label{eq:La2alpha}
  \Lambda_2(\alpha) &=& \frac{1}{1 + (k_2/k_1)(\beta_1/\beta_2)\sqrt{\mu_1 \alpha/\mu_2}}\\
  &\approx& \frac{1}{1 + (k_2/k_1)(m_1/m_2)\sqrt{\alpha}}, \nonumber
\end{eqnarray}
At the exact commensurability, we have
\begin{equation}
  \label{eq:alpha0}
  \alpha_0=\left(\frac{\mu_2}{\mu_1}\right)^{1/3}\left(\frac{k_1}{k_2}\right)^{2/3}
  \approx \left(\frac{k_1}{k_2}\right)^{2/3}
\end{equation}
The quantities $G_i$ depend on $\alpha$ and on the planet eccentricities
\begin{equation}
  \label{eq:Gie}
  G_i(\alpha,e_i) = \Lambda_i(\alpha) \sqrt{1-e_i^2}.
\end{equation}
Let us note $I_i$ the renormalized angular momentum deficit (AMD) of planet $i$
\citep{laskar_large_1997,laskar_spacing_2000}
\begin{equation}
  \label{eq:Di}
  I_i = \Lambda_i - G_i
  = \frac{1}{2} \Lambda_i \xi_i^2 \propto e_i^2,
\end{equation}
with
\begin{equation}
  \label{eq:xi}
  \xi_i = \sqrt{2\left(1-\sqrt{1-e_i^2}\right)} \approx e_i.
\end{equation}
The simplifying assumption introduced in \citet{delisle_resonance_2014} implies
(see also Appendix~\ref{sec:simplif})
\begin{equation}
  \label{eq:assum}
  \frac{I_2}{I_1} = \frac{I_{2,ell}}{I_{1,ell}} \equiv \tan^2\phi,
\end{equation}
where $\phi$ is a constant angle and $I_{i,ell}$ are values of the renormalized AMD at the center
of the resonance \citep[elliptical fixed point, see][]{delisle_resonance_2014}.
We also note $\mathcal{D}$ the renormalized total AMD
\begin{equation}
  \label{eq:AMD}
  \mathcal{D} = I_1 + I_2 = \delta+\epsilon.
\end{equation}
The parameter $\delta$ corresponds to the renormalized total AMD at the exact commensurability
($\delta = \mathcal{D}_0$).
Thus, for a resonant system, $\delta$ provides a measure of the planet eccentricities
($\delta \propto e^2$, see Eq.~(\ref{eq:Di})).
Figure~\ref{fig:I} shows the phase space corresponding
to Hamiltonian~(\ref{eq:Hqeps}).
The width of the resonant area is proportional to $\delta^{q/4} \propto e^{q/2}$
for a resonance of order $q$ (see Fig.~\ref{fig:I}).
For a resonant system, in the regime of moderate eccentricities,
a measure (between 0 and 1) of the relative amplitude of libration
(amplitude of libration versus resonance width) is given by
\citep[see][]{delisle_resonance_2014}
\begin{equation}
  \label{eq:Ampli}
  A  = \sin^2\left(\frac{q\theta_{max}}{2}\right),
\end{equation}
where $\theta_{max}$ is the maximum value reached by the resonant angle $\theta$ during a libration period (see Fig.~\ref{fig:I}).

\figI

Note that our simplifying assumption
(eccentricity ratio close to the forced eccentricities ratio)
is well verified when the amplitude of libration is small ($A \ll 1$)
and the system stays close to the elliptical fixed point.
For high amplitude of libration ($A \sim 1$), the eccentricity ratio undergoes
oscillations around the forced value and our model only provide a first approximation
of the motion \citep[see][]{delisle_resonance_2014}.

\section{Resonant motion in the dissipative case}
\label{sec:dissip}

In this section we describe the evolution of a resonant system undergoing dissipation.
The main parameters that have to be tracked during this evolution are the parameter $\delta$
which describes the evolution
of the phase space (and of the eccentricities for resonant systems)
and the relative amplitude $A$ which describes the spiraling of the trajectory
with respect to the separatrix of the resonance.

\subsection{General case}
\label{sec:general}

Let us consider a dissipative force acting on the semi-major axes and the eccentricities of both planets.
We first consider a very general case and do not assume a particular form for this dissipation, except that it
acts on a long timescale.
The evolution of the system can be described by the three following timescales
(which may depend on the eccentricities and semi-major axes of the planets):
$(\xi_1/\dot{\xi}_1)_d$, $(\xi_2/\dot{\xi}_2)_d$, $(\alpha/\dot{\alpha})_d$.
Note that, for sufficiently small eccentricities, we have $\xi_i\approx e_i$, and
\begin{equation}
  \label{eq:dxidt}
  \left.\frac{\dot{\xi}_i}{\xi_i}\right|_d \approx  \left.\frac{\dot{e}_i}{e_i}\right|_d.
\end{equation}

The evolution of the parameter $\delta$ that drives the evolution of the phase space
(and of the eccentricities for resonant systems) is given by (see Appendix~\ref{sec:evo-delta})
\begin{eqnarray}
  \label{eq:ddeltadt}
  \dot{\delta}|_d
  &=& 2 \left(\cos^2\phi \left.\frac{\dot{\xi}_1}{\xi_1}\right|_d
    + \sin^2\phi \left.\frac{\dot{\xi}_2}{\xi_2}\right|_d \right) \mathcal{D}\nonumber \\
  && + \frac{\Lambda_2 - \sin^2\phi}{2} \left.\frac{\dot{\alpha}}{\alpha}\right|_d \mathcal{D}\\
  && + \frac{q}{k_1} \frac{\Lambda_1\Lambda_2}{2} \left.\frac{\dot{\alpha}}{\alpha}\right|_d.\nonumber
\end{eqnarray}
For a resonant system, the evolution of the relative amplitude of libration reads
\citep[see][Appendix A]{delisle_resonance_2014}
\begin{equation}
  \label{eq:dAdt}
  <\dot{A}> = \frac{1}{2 R \delta^{q/2}} \left( <\epsilon \dot{\epsilon}|_d>
    - \frac{q}{4\delta}<\epsilon^2\dot{\delta}|_d> \right),
\end{equation}
with
\begin{equation}
  \dot{\epsilon}|_d = - \frac{q}{k_1} \frac{\Lambda_1\Lambda_2}{2} \left.\frac{\dot{\alpha}}{\alpha}\right|_d.
\end{equation}

\subsection{Disk-planet interactions}
\label{sec:disk}

Let us now apply Eqs.~(\ref{eq:ddeltadt}),~(\ref{eq:dAdt}) to the specific
case of disk-planet interactions.
Because of these interactions the planets undergo a
torque that induces a modification in their orbital elements and
subsequent migration in the disk
\citep[e.g.][]{goldreich_excitation_1979,goldreich_disk_1980}.
In particular, the angular momentum of each planet evolves on an exponential timescale
$T_{m,i}$ due to this migration,
while eccentricities evolve on an
exponential timescale $T_{e,i}$
\citep[e.g.][]{papaloizou_orbital_2000,terquem_migration_2007,goldreich_overstable_2014}:
\begin{eqnarray}
  \left.\frac{\dot{\hat{G}}_i}{\hat{G}_i}\right|_d &=& - \frac{1}{T_{m,i}},\\
  \left.\frac{\dot{e}_i}{e_i}\right|_d &=& - \frac{1}{T_{e,i}},
\end{eqnarray}
where $\hat{G}_i$ is the angular momentum of planet $i$.
From these simple decay laws we can deduce the evolution of the parameters of interest
for resonant systems ($\dot{\delta}$ and $\dot{A}$, see Sect.~\ref{sec:general}).
We have
\begin{equation}
  \label{eq:diskTe}
  \left.\frac{\dot{\xi}_i}{\xi_i}\right|_d \approx \left.\frac{\dot{e}_i}{e_i}\right|_d = -\frac{1}{T_{e,i}},
\end{equation}
\begin{equation}
  \label{eq:diskTa}
  \left.\frac{\dot{a}_i}{a_i}\right|_d = -\frac{2}{T_{m,i}} + 2  \frac{\xi_i^2}{1-\xi_i^2} \left.\frac{\dot{\xi}_i}{\xi_i}\right|_d
  \approx -\frac{2}{T_{m,i}} - 2 \frac{\xi_i^2}{T_{e,i}}.
\end{equation}
The evolution of the semi-major axis ratio is thus governed by
\begin{equation}
  \label{eq:diskTalpha}
  \left.\frac{\dot{\alpha}}{\alpha}\right|_d
  = \frac{2}{T_m} + \frac{4}{\Lambda_1\Lambda_2} \left(\frac{\Lambda_1\sin^2\phi}{T_{e,2}} - \frac{\Lambda_2\cos^2\phi}{T_{e,1}} \right) \mathcal{D},
\end{equation}
with
\begin{equation}
  \label{eq:diskTmig}
  \frac{1}{T_m} = \frac{1}{T_{m,2}} - \frac{1}{T_{m,1}}.
\end{equation}
From Eq.~(\ref{eq:ddeltadt}) we obtain
\begin{eqnarray}
  \dot{\delta}|_d &=&
  \frac{q}{k_1} \frac{\Lambda_1\Lambda_2}{T_m}\nonumber\\
  && + \frac{\Lambda_2-\sin^2\phi}{T_m} \mathcal{D} \nonumber\\
  && - 2 \left(\frac{\cos^2\phi}{T_{e,1}}+\frac{\sin^2\phi}{T_{e,2}}\right) \mathcal{D} \nonumber\\
  && + 2 \frac{q}{k_1} \left(\frac{\Lambda_1\sin^2\phi}{T_{e,2}} - \frac{\Lambda_1\cos^2\phi}{T_{e,1}} \right) \mathcal{D}
  \label{eq:diskdelta}\\
  &=& \frac{q}{k_1}\frac{\Lambda_1\Lambda_2}{T_m} \nonumber\\
  && - \mathcal{D} \left[ 2 (\Lambda_1+\Lambda_2) \left( \frac{k_2}{k_1} \frac{\cos^2\phi}{T_{e,1}} + \frac{\sin^2\phi}{T_{e,2}} \right)\right.\nonumber\\
  && \qquad \left.  - \frac{\Lambda_2-\sin^2\phi}{T_m}\right] ,\nonumber
\end{eqnarray}
where we neglect second order terms in $\mathcal{D}$ ($\mathcal{D}^2 \propto e^4$).
Let us note
\begin{eqnarray}
  \frac{1}{T_M} &=& \frac{q}{k_1}\frac{\Lambda_1\Lambda_2}{T_m},\\
  \frac{1}{T_E} &=&  2 (\Lambda_1+\Lambda_2) \left( \frac{k_2}{k_1} \frac{\cos^2\phi}{T_{e,1}} + \frac{\sin^2\phi}{T_{e,2}} \right)
  - \frac{\Lambda_2-\sin^2\phi}{T_m}.
\end{eqnarray}
We thus have
\begin{eqnarray}
  \dot{\delta}|_d &=& \frac{1}{T_M} - \frac{\mathcal{D}}{T_E},\\
  \label{eq:diskdeltamean}
  <\dot{\delta}|_d> &=& \frac{1}{T_M} - \frac{\delta}{T_E}.
\end{eqnarray}
Note that the damping timescale is often much shorter than the migration timescale \citep[$T_{e,i}\ll T_{m,i}$, e.g.][]{goldreich_disk_1980},
thus
\begin{equation}
  \label{eq:TEapprox}
  \frac{1}{T_E} \approx  2 (\Lambda_1+\Lambda_2) \left( \frac{k_2}{k_1} \frac{\cos^2\phi}{T_{e,1}} + \frac{\sin^2\phi}{T_{e,2}} \right).
\end{equation}
The timescales $T_E$, $T_M$ can be expressed using more usual notations
\begin{eqnarray}
  \label{eq:TM}
  \frac{1}{T_M} &\approx& \frac{q}{k_1}
  \left(2\frac{k_2}{k_1} + \frac{m_2}{m_1}\frac{1}{\sqrt{\alpha_0}} +
    \left(\frac{k_1}{k_2}\right)^2\frac{m_1}{m_2}\sqrt{\alpha_0}\right)^{-1} \times\\
  && \left(\frac{1}{T_{m,2}} - \frac{1}{T_{m,1}}\right),\nonumber\\
  \label{eq:TE}
  \frac{1}{T_E} &\approx& 2 \left(1+\frac{m_1}{m_2}\sqrt{\alpha_0}\right)
  \left(1+\frac{k_2}{k_1}\frac{m_1}{m_2}\sqrt{\alpha_0}\right)^{-1}\times \\
  && \left(\frac{1}{T_{e,2}}+\frac{k_2}{k_1}\frac{m_1}{m_2}\left(\frac{e_1}{e_2}\right)_{ell}^2\hspace{-2mm}\sqrt{\alpha_0}\frac{1}{T_{e,1}}\right)
  \left(1+\frac{m_1}{m_2}\left(\frac{e_1}{e_2}\right)_{ell}^2\hspace{-2mm}\sqrt{\alpha_0}\right)^{-1}.\nonumber
\end{eqnarray}

Depending on the values of $T_M$ and $T_E$, different evolution scenarios for $\delta$ are possible.
Note that all these equations remain valid for $T_M$ and $T_E$ negative.
In most cases the disk induces a damping of eccentricities ($T_{e,i} > 0$, thus $T_E>0$),
but some studies \citep[e.g.][]{goldreich_eccentricity_2003} suggest
that an excitation of the eccentricities by the disk is possible ($T_{e,i} < 0$, thus $T_E<0$).
$T_M$ is positive if the period ratio between the planets ($P_2/P_1$) decreases (convergent migration).
But if the planets undergo divergent migration ($P_2/P_1$ increases), $T_M$ is negative.
This does not depends on the absolute direction (inward or outward)
of the migration of the planets in the disk but only on the evolution of their period ratio.

In the case of divergent migration, the planets cannot get trapped in resonance \citep[e.g.][]{henrard_second_1983}.
The system always ends-up with a period ratio higher than the resonant value and this does not depend
on the damping/excitation of eccentricities.

The case of convergent migration is more interesting.
If the initial period ratio is higher than the resonant value, the planets can be locked in resonance.
This induces an excitation of the eccentricities of the planets ($\dot\delta|_M = 1/T_M> 0$).
If $T_E<0$ (excitation of eccentricities by the disk) or $T_E \gg T_M$ (inefficient damping),
$\delta$ (as well as the eccentricities) does not stop increasing.
When eccentricities reach too high values,
the system becomes unstable and the resonant configuration is broken.

The most common scenario is the case of efficient damping of eccentricities ($0 < T_E \lesssim T_M$).
In this case, $\delta$ reaches an equilibrium value ($<\dot\delta|_d> = 0$, see Eq.~(\ref{eq:diskdeltamean}))
\begin{eqnarray}
  \delta_{eq} &=& \frac{T_E}{T_M}\nonumber\\
  &=& \frac{q}{2 k_1} \frac{\Lambda_1\Lambda_2}{\Lambda_1+\Lambda_2}
  \left(\frac{1}{T_{m,2}}-\frac{1}{T_{m,1}}\right)
  \left(\frac{k_2}{k_1}\frac{\cos^2\phi}{T_{e,1}}+\frac{\sin^2\phi}{T_{e,2}}\right)^{-1}\nonumber\\
  \label{eq:diskdeltaeq}
  &\approx& \frac{q}{2}
  \left(1+\frac{m_1}{m_2}\left(\frac{e_1}{e_2}\right)_{ell}^2\hspace{-2mm}\sqrt{\alpha_0}\right)
  \times\\
  && \left(k_1+k_2 + k_2\frac{m_1}{m_2}\sqrt{\alpha_0} +
    k_1\frac{m_2}{m_1}\frac{1}{\sqrt{\alpha_0}}\right)^{-1}
  \times \nonumber\\
  && \left(\frac{1}{T_{m,2}}-\frac{1}{T_{m,1}}\right)
  \left(\frac{1}{T_{e,2}}+\frac{k_2}{k_1}\frac{m_1}{m_2}\left(\frac{e_1}{e_2}\right)_{ell}^2\hspace{-2mm}\sqrt{\alpha_0}\frac{1}{T_{e,1}}\right)^{-1}.\nonumber
\end{eqnarray}
However, as shown by \citet{goldreich_overstable_2014} for the restricted three body problem,
this equilibrium can be unstable.
Even if the parameter $\delta$ reaches the equilibrium $\delta_{eq}$
and the phase space of the system stops evolving,
the amplitude of libration can increase until the system crosses the separatrix and escapes from resonance.

Let us now compute the evolution of this amplitude of libration.
According to Eq.~(\ref{eq:dAdt}), we need to compute
$<\epsilon \dot{\epsilon}|_d>$ and  $<\epsilon^2\dot{\delta}|_d>$.
We have
\begin{eqnarray}
  \label{eq:diskA1}
  <\epsilon \dot{\epsilon}|_d> &=&
  2 \frac{q}{k_1} \left(\frac{\Lambda_2\cos^2\phi}{T_{e,1}} -\frac{\Lambda_1\sin^2\phi}{T_{e,2}}\right) <\epsilon^2>,\\
  \label{eq:diskA2}
  <\epsilon^2\dot{\delta}|_d> &=& \left(\frac{1}{T_M} - \frac{\delta}{T_E} \right) <\epsilon^2>
  = <\dot{\delta}|_d><\epsilon^2>,
\end{eqnarray}
where $<\epsilon^2>$ can be computed using elliptic integrals \citep[see][]{delisle_resonance_2014}
\begin{equation}
  \label{eq:tidePlaA3}
  <\epsilon^2> \approx 2 R \delta^{q/2} A.
\end{equation}

Note that the first term ($<\epsilon \dot{\epsilon}|_d>$) does not depend on the migration timescale but only on the damping timescale.
The second term ($<\epsilon^2\dot{\delta}|_d>$) vanishes
when the system reaches the equilibrium $\delta = \delta_{eq}$, since $\dot\delta|_d=0$.
This is not surprising because the first term describes the evolution of the absolute amplitude of libration
$\epsilon^2$, while the second one describes the evolution of the resonance width which does not evolve if
the phase space does not evolve (constant $\delta$).
Finally, we obtain (see Eq.~(\ref{eq:dAdt}))
\begin{equation}
  \label{eq:diskAtot}
  \left.\frac{\dot{A}}{A}\right|_d =
  2 \frac{q}{k_1} \left(\frac{\Lambda_2\cos^2\phi}{T_{e,1}} - \frac{\Lambda_1\sin^2\phi}{T_{e,2}}\right).
\end{equation}
The amplitude of libration increases if
\begin{equation}
  \frac{\Lambda_2\cos^2\phi}{T_{e,1}} > \frac{\Lambda_1\sin^2\phi}{T_{e,2}},
\end{equation}
which is equivalent to
\begin{equation}
  \label{eq:diskAincr}
  \frac{T_{e,1}}{T_{e,2}} < \frac{}{}\left(\frac{\xi_1}{\xi_2}\right)_{ell}^2.
\end{equation}
Using $\xi_i\approx e_i$, this gives
\begin{equation}
  \label{eq:diskCritA}
  A \nearrow  \qquad \Longleftrightarrow \qquad \frac{T_{e,1}}{T_{e,2}} < \left(\frac{e_1}{e_2}\right)_{ell}^2,
\end{equation}
where the eccentricity ratio is evaluated at the elliptical fixed point ($ell$ subscript) at the center of the resonance.
In the circular restricted case studied by \citet{goldreich_overstable_2014},
$e_2 = 0$ and $T_{e,2} = +\infty$, thus the amplitude always increases (Eq.~(\ref{eq:diskCritA}))
and the equilibrium is unstable.
However, in the opposite restricted case ($e_1 = 0$ and $T_{e,1} = +\infty$),
that was not addressed in \citet{goldreich_overstable_2014}, the amplitude always decreases (Eq.~(\ref{eq:diskCritA})) leading to a stable equilibrium.

Note that this result is based on our approximation that the eccentricity ratio
remains close to the forced value (value at the elliptical fixed point).
This is well verified for a small amplitude of libration but when the
amplitude increases, the eccentricity ratio oscillates and may differ significantly from the
forced value. Our model thus only provides a first approximation of the mean value of this ratio
in the case of high amplitude of libration.

To sum up, the evolution of a resonant pair of planets undergoing disk-planet interactions depends
mainly on two parameters:
$T_E/T_M$ (damping vs migration timescale)
and $T_{e,1}/T_{e,2}$ (damping in inner planet vs outer planet).
The ratio $T_E/T_M$ governs the equilibrium eccentricities of the planets
(see Eqs.~(\ref{eq:diskdeltamean}), (\ref{eq:diskdeltaeq})).
The ratio $T_{e,1}/T_{e,2}$ governs the stability of this equilibrium
(see Eqs.~(\ref{eq:diskAtot}), (\ref{eq:diskCritA})).

\section{Constraints on disk properties}
\label{sec:applications}

In this section, we show how the classification of the outcome of disk-planet interactions can be used to put
constraints on the dissipative forces undergone by the planets and thus on some disk properties.
More precisely, if a system is currently observed to harbour two planets locked in a MMR,
it is probable that this resonant configuration was stable (or unstable but with a very long timescale) when the disk was present.
We could imagine that the configuration was highly unstable but the protoplanetary disk disappeared before the
system had time to escape from resonance, however this would require a fine tuning of the disk disappearing timing.
Thus the amplitude of libration was probably either decreasing, or increasing on a very long timescale.
This induces that
\begin{equation}
  \frac{T_{e,1}}{T_{e,2}} \gtrsim \left(\frac{e_1}{e_2}\right)_{ell}^2.
\end{equation}

Moreover, a small amplitude of libration is probably the sign of a damping of the amplitude on a short timescale
\begin{equation}
  \label{eq:constSmallA}
  \frac{T_{e,1}}{T_{e,2}} \gg \left(\frac{e_1}{e_2}\right)_{ell}^2.
\end{equation}
On the opposite, a large amplitude of libration could be the sign of
a long timescale of amplitude damping or a long timescale of amplitude excitation.
Indeed, if the amplitude was increasing fast,
the system should not be observed in resonance.
If it was decreasing fast, the observed amplitude should be very small.
However, another mechanism may be responsible for the excitation of the amplitude of libration, possibly
after the disk disappearing (e.g. presence of a third planet in the system).
Thus we cannot exclude the case of a fast damping of the amplitude of libration, even
in the case of an observed large amplitude,
\begin{equation}
  \label{eq:constBigA}
  \frac{T_{e,1}}{T_{e,2}} \gtrsim \left(\frac{e_1}{e_2}\right)_{ell}^2.
\end{equation}

In addition to the constraints obtained from the observed amplitude of libration,
the observed values of both eccentricities is also an important information.
If the planets did not undergo other source of dissipation since the disk has disappeared,
the present eccentricities should still correspond to the equilibrium ones.
For close-in planets, the tides raised by the star on the planets induce a significant
dissipative evolution of the system after the disk disappearing \citep[e.g.][]{delisle_tidal_2014}.
Therefore, this reasoning only applies for systems farther from the star for which tidal interactions have a negligible
effect on the orbits over the age of the system.
Let us recall that the equilibrium eccentricities are given by (Eq.~(\ref{eq:diskdeltaeq}))
\begin{equation}
  \delta = \delta_{eq} = \frac{T_E}{T_M},
\end{equation}
with
\begin{eqnarray}
  \delta &\approx& \frac{1}{2} (\Lambda_1 \xi_1^2 + \Lambda_2 \xi_2^2)\nonumber\\
  \label{eq:deltaestim}
  &\approx& \left(\frac{k_2}{k_1} + \frac{m_2}{m_1}\frac{1}{\sqrt{\alpha}}\right)^{-1} \frac{e_1^2}{2}
  + \left(1 + \frac{k_2}{k_1}\frac{m_1}{m_2}\sqrt{\alpha}\right)^{-1} \frac{e_2^2}{2}.
\end{eqnarray}
$\delta$ can be computed from the known (observed) orbital elements of the planets.
Thus, the ratio $T_E/T_M$ of the damping and migration timescales is constrained by the observations.
This ratio depends on the four timescales ($T_{e,1}$, $T_{e,2}$, $T_{m,1}$, and $T_{m,2}$) of the model (see Eq.~(\ref{eq:diskdeltaeq})) which themselves
depend on the properties of the disk and the planets.
There exists a wide diversity of disk models in the literature,
which would result in significantly different
migration and damping timescales for each planet.
Our analytical model is very general and can handle these different models as long as
expressions for $T_{e,1}$, $T_{e,2}$, $T_{m,1}$, and $T_{m,2}$ are available.

We consider here the case of type I migration to illustrate
the possibility of constraining the disk properties for observed systems.
Following the prescriptions of
\citet{kley_planet_2012}, we have
\begin{eqnarray}
  \frac{T_{m,1}}{T_{m,2}} &\approx& \frac{m_2}{m_1} \sqrt{\frac{a_2}{a_1}} \left(\frac{H(a_1)/a_1}{H(a_2)/a_2}\right)^2 \frac{\Sigma(a_2)}{\Sigma(a_1)},\\
  \frac{T_{e,i}}{T_{m,i}} &\approx& \left(\frac{H(a_i)}{a_i}\right)^2,
\end{eqnarray}
where $H(a)$ is the local disk scale height and
$\Sigma(a) \propto a^{-\beta_\Sigma}$ its local surface density.
The standard MMSN (Minimum Mass Solar Nebula) model assumes
$\Sigma \propto a^{-3/2}$ ($\beta_{\Sigma} = 3/2$).
$H/a$ is called the disk aspect ratio and is often assumed to be roughly constant and of the
order of 0.05 \citep[e.g.][]{kley_planet_2012}.
Using these assumptions we obtain
\begin{eqnarray}
  \frac{T_{m,1}}{T_{m,2}} &\approx& \frac{m_2}{m_1} \alpha^{\beta_{\Sigma}-1/2},\\
  \frac{T_{m,1}}{T_{e,1}} &\approx& \frac{T_{m,2}}{T_{e,2}} \approx \left(\frac{H}{a}\right)^{-2}.
\end{eqnarray}
For the sake of brevity, we note in the following
\begin{eqnarray}
  \label{eq:deftau}
  \tau &=&  \frac{m_2}{m_1} \alpha^{\beta_{\Sigma}-1/2} \approx \frac{T_{m,1}}{T_{m,2}} \approx \frac{T_{e,1}}{T_{e,2}}, \\
  \label{eq:defK}
  K &=& \left(\frac{H}{a}\right)^{-2} \approx \frac{T_{m,1}}{T_{e,1}} \approx \frac{T_{m,2}}{T_{e,2}}.
\end{eqnarray}
If the system is observed with a small amplitude of libration in the resonance,
we have (Eq.~(\ref{eq:constSmallA}))
\begin{equation}
  \label{eq:constSmallAbis}
  \tau \gg \left(\frac{e_1}{e_2}\right)_{ell}^2,
\end{equation}
and if the amplitude is large we have (Eq.~(\ref{eq:constBigA}))
\begin{equation}
  \label{eq:constBigAbis}
  \tau \gtrsim \left(\frac{e_1}{e_2}\right)_{ell}^2.
\end{equation}

The lower limit we obtain for $\tau$ corresponds to an upper limit for the density profile exponent
$\beta_\Sigma$ (see Eq.~(\ref{eq:deftau})).
In particular, if
\begin{equation}
  \label{eq:InvertedProfile}
  \frac{m_2}{m_1}\alpha^{-1/2} < \left(\frac{e_1}{e_2}\right)_{ell}^2,
\end{equation}
the density profile of the disk must be inverted
($\beta_\Sigma<0$, i.e. the surface density increases with the distance to the star)
for the system to be stable in resonance.
The condition of Eq.~(\ref{eq:InvertedProfile}) is roughly equivalent to
\begin{equation}
  I_1 > I_2,
\end{equation}
where $I_1$, $I_2$ are the angular momentum deficits (AMD) of both planets.

Let us recall that in order to be captured in resonance
the planets must undergo convergent migration \citep[e.g.][]{henrard_second_1983}.
This put another constraint on the parameter $\tau$ (see Eq.~(\ref{eq:deftau}))
\begin{equation}
  \label{eq:constMigConv}
  \tau > 1.
\end{equation}
Again, this corresponds to an upper limit for $\beta_\Sigma$, and if
\begin{equation}
  \label{eq:InvertedProfilebis}
  \frac{m_2}{m_1}\alpha^{-1/2} < 1,
\end{equation}
the density profile of the disk must be inverted ($\beta_\Sigma<0$)
for the planets to undergo convergent migration.
The condition of Eq.~(\ref{eq:InvertedProfilebis}) is roughly equivalent to
\begin{equation}
  \Lambda_1 > \Lambda_2,
\end{equation}
where $\Lambda_1$, $\Lambda_2$ are the circular angular momenta of both planets.

The constraint provided by the observation of the equilibrium eccentricities reads
(see Eq.~(\ref{eq:diskdeltaeq}))
\begin{equation}
  \delta_{eq} = \frac{C_1}{K} \frac{\tau-1}{\tau + C_2},
\end{equation}
with
\begin{eqnarray}
  \label{eq:C1}
  C_1 &=& \frac{1}{2}
  \left(1+\frac{m_1}{m_2}\left(\frac{e_1}{e_2}\right)_{ell}^2\hspace{-2mm}\sqrt{\alpha_0}\right)
  \times\\
  && \left(1 + \frac{k_2}{q}\frac{m_1}{m_2}\sqrt{\alpha_0} +
    \frac{k_1}{q}\frac{m_2}{m_1}\frac{1}{\sqrt{\alpha_0}}\right)^{-1},
  \nonumber\\
  \label{eq:C2}
  C_2 &=& \frac{k_2}{k_1}\frac{m_1}{m_2}\left(\frac{e_1}{e_2}\right)_{ell}^2\hspace{-2mm}\sqrt{\alpha_0},
\end{eqnarray}
and $\delta_{eq}$ is given by Eq.~(\ref{eq:deltaestim}).
We thus have
\begin{equation}
  \label{eq:constK}
  K = \frac{C_1}{\delta} \frac{\tau-1}{\tau + C_2},
\end{equation}
where $C_1$, $C_2$ and $\delta$ can all be derived from the observations.
Note that $K$ is an increasing function of $\tau$ (Eq.~(\ref{eq:constK})).
Thus, our analytical criterion for stability provides a lower bound for both $\tau$ and $K$.

\section{Application to observed resonant systems}
\label{sec:appli}

In the following we apply our analytical criteria to systems that are observed in resonance.
We also performed N-body simulations with the additional migration and damping forces
exerted by the disk on the planets.
We used the ODEX integrator \citep[e.g][]{hairer_solving_2010}
and the dissipative timescales $T_{m,i}$ (angular momentum evolution),
$T_{e,i}$ (eccentricity evolution) are fixed for each planet and each simulation.

Many multi-planetary systems are observed close to different MMR
(period ratio close to a resonant value).
However only a few of them have a precise enough determination of the planets
orbital parameters in order
to distinguish between resonant motion and near-resonant motion.
To our knowledge, all known resonant planet pairs are giant planets
(better precision of orbital parameters).
We thus selected three of these resonant giant planet pairs to illustrate our model.

Note that giant planets are believed to undergo type II migration.
Our analytical model is very general and can take into account any prescription
for the evolution of the angular momentum and the eccentricity of each planet.
We did not find a simple analytical prescription for type II migration
in the literature.
Indeed type II migration is a more complex (non-linear) mechanism than type I migration,
and it is not yet well understood.
In particular, the timescale of type II migration is still discussed \citep[e.g.][]{duffell_migration_2014,durmann_migration_2015}.
However, the effect of type II migration is expected to be similar to type I migration
\citep[i.e. inward migration and damping of eccentricities, e.g.][]{bitsch_orbital_2010}.
The main difference is that the disk profile is affected by the presence of the planets
(gap around the planets) and the timescales of migration and damping are thus slowed down.
We thus chose to apply type I migration prescriptions for our study
of these giant planets resonant systems as a first approximation.

\subsection{HD~60532b, c: 3:1 MMR, large amplitude of libration}
\label{sec:hd60532}

\tabI

\figII

The star \object{HD~60532} hosts two planets \citep[see][]{desort_extrasolar_2008} that
exhibit a 3:1 period ratio.
\citet{laskar_planetary_2009} performed a dynamical study of the system and confirmed the
3:1 MMR between the planets with a large amplitude of libration ($\sim 40^\circ$).
We reproduced the orbital elements of the planets
\citep[taken from][]{laskar_planetary_2009} in Table~\ref{tab:I}.
For this system, the forced eccentricity ratio (ratio of eccentricity at the center of the resonance) is
\begin{equation}
 \left(\frac{e_1}{e_2}\right)_{ell} \approx 3.
\end{equation}
Since the system is observed with a large amplitude of libration, the stability constraint
gives (Eq.~(\ref{eq:constBigAbis}))
\begin{equation}
  \tau \gtrsim  \left(\frac{e_1}{e_2}\right)_{ell}^2 \approx 9.
\end{equation}
Note that this value is much larger than one, thus the condition of convergent
migration is fulfilled (see Eq.~(\ref{eq:constMigConv})).
This value of $\tau$ corresponds to a surface density profile exponent of
(see Eq.~(\ref{eq:deftau}))
\begin{equation}
  \beta_\Sigma \lesssim -1.3.
\end{equation}
Let us recall that for the MMSN model $\beta_\Sigma = 3/2$.
The negative value we obtain corresponds to an inverted density profile.
This inverted density profile is very untypical of type I migration.
This result is thus an evidence that the planets did not undergo a classical type I migration.
This is not surprising since giant planets are expected to open a gap and undergo type II migration.
Our results also constrain a type II migration scenario.
Indeed, independently of the migration prescriptions,
for the resonance to be stable the damping of the outer eccentricity must be
much more efficient than the inner eccentricity damping ($T_{e,1}/T_{e,2} \gtrsim 9$).
One would need prescriptions for type II migration to relate this result
with some disk and/or planets properties.

From the observed orbital elements we obtain
(see Eqs.~(\ref{eq:deltaestim}), (\ref{eq:C1}), and (\ref{eq:C2}))
\begin{eqnarray}
  \delta &=& 6.5\times 10^{-3},\\
  C_1 &=& 0.44,\\
  C_2 &=& 7.9.
\end{eqnarray}
Using $\tau\gtrsim 9$, the current eccentricities should be reproduced with
(see Eq.~(\ref{eq:constK}))
\begin{equation}
  K \gtrsim 30,
\end{equation}
which corresponds to an aspect ratio of (see Eq.~(\ref{eq:defK}))
\begin{equation}
  \frac{H}{a} \lesssim 0.18.
\end{equation}

We performed numerical simulations with different values of $\tau$.
For each simulation, the value of $K$ is computed using Eq.~(\ref{eq:constK}),
in order to reproduce the equilibrium eccentricities.
We fixed $T_{m,2} = 5\times 10^5$~yr for all simulations,
and integrated the system for $10^6$~yr.
We thus have
$T_{m,1} = 5\times 10^5 \tau$~yr,
$T_{e,2} = 5\times 10^5/K$~yr,
$T_{e,1} = 5\times 10^5 \tau/K$~yr.
The semi-major axes are initially 10 and 22~AU, such that the system is initially outside
of the 3:1 resonance with a period ratio of about 3.3.
Both eccentricities are initially set to 0.001 with anti-aligned periastrons and
coplanar orbits.
The planets are initially at periastrons (zero anomalies).
The evolution of the semi-major axes, the period ratio, the eccentricities,
the eccentricity ratio, and the angles are shown in Fig.~\ref{fig:II}.
These simulations confirm our analytical results: for $\tau<9$, the amplitude
of libration decreases, and for $\tau>9$ the amplitude increases (see Fig.~\ref{fig:II}).

\subsection{GJ~876b, c: 2:1 MMR, small amplitude of libration}
\label{sec:gj876}

\tabII

\figIII

\object{GJ~876} is a M-dwarf hosting four planets
\citep{delfosse_closest_1998,marcy_planetary_1998,marcy_pair_2001,rivera_lick_2010}.
The planet b and c, in which we are interested here, are Jupiter-mass planets embedded
in a 2:1 MMR, while d and e are much less massive.
A small amplitude of libration is observed
\citep[$\sim 5\degr$, see][]{correia_harps_2010}
for the 2:1 resonance between \object{GJ~876}b, c.
The forced eccentricity ratio is
\begin{equation}
 \left(\frac{e_1}{e_2}\right)_{ell} \approx 6.5.
\end{equation}
Since the system is observed with a small amplitude of libration, the stability constraint
gives (Eq.~(\ref{eq:constSmallAbis}))
\begin{equation}
  \tau \gg \left(\frac{e_1}{e_2}\right)_{ell}^2 \approx 42.
\end{equation}
As for \object{HD~60532}b, c, the condition of convergent
migration is fulfilled (see Eq.~(\ref{eq:constMigConv})).
The surface density profile exponent is  (see Eq.~(\ref{eq:deftau}))
\begin{equation}
  \beta_\Sigma \ll -5.2.
\end{equation}
We obtain again a negative value that correspond to an inverted profile.
We have
(see Eqs.~(\ref{eq:deltaestim}), (\ref{eq:C1}), and (\ref{eq:C2}))
\begin{eqnarray}
  \delta &=& 6.4\times 10^{-3},\\
  C_1 &=& 0.81,\\
  C_2 &=& 22.
\end{eqnarray}
Using these values and $\tau \gg 42$, we obtain
(see Eq.~(\ref{eq:constK}))
\begin{equation}
  K \gg 80,
\end{equation}
which corresponds to an aspect ratio of (see Eq.~(\ref{eq:defK}))
\begin{equation}
  \frac{H}{a} \ll 0.11.
\end{equation}

As for \object{HD~60532}b, c we performed numerical simulations
with different values of $\tau$
(and adjusted values of $K$ given by Eq.~(\ref{eq:constK})).
The semi-major axes are initially 2 and 3.5~AU (period ratio of about 2.3),
the eccentricities are 0.001 with anti-aligned periastrons and
coplanar orbits.
The planets are initially at periastrons (zero anomalies).
The evolution of the semi-major axes, the period ratio, the eccentricities,
the eccentricity ratio, and the angles are shown in Fig.~\ref{fig:III}.
The transition between decreasing and increasing amplitude of libration
happens around $\tau\approx 20$ (see Fig.~\ref{fig:III}),
while our analytical criterion gives a value of 42.
Taking this refined value for $\tau$, the condition for reproducing the
observed system with type I migration reads
\begin{eqnarray}
  \beta_\Sigma &\ll& -3.5,\\
  K &\gg& 58,\\
  \frac{H}{a} &\ll& 0.13.
\end{eqnarray}
The density profile still needs to be inverted ($\beta_\Sigma<0$) and the
overall conclusions are the same.

Note that \citet{lee_dynamics_2002} studied capture scenarios for this system using a slightly different model
for the migration and damping and did not observe any evolution of the libration amplitude in their simulations.
The authors used constant semi-major axis ($T_{a,i}$) and eccentricity ($T_{e,i}$) damping timescales for each planet.
In our study we followed the prescriptions of \citet{papaloizou_orbital_2000} \citep[see also][]{goldreich_overstable_2014} and
considered constant angular momentum ($T_{m,i}$) and eccentricity ($T_{e,i}$) damping timescales.
We replaced these prescriptions with \citet{lee_dynamics_2002} prescriptions for the disk-planet interactions in our analytical model
(following the same scheme as described in Sect.~\ref{sec:disk}).
We found that the amplitude of libration does not evolve in this case
(in agreement with \citealt{lee_dynamics_2002} simulations).
This difference between both prescriptions has important consequences since
in the case of \citet{lee_dynamics_2002} prescriptions two initially resonant
planets will stay locked in resonance forever while
with the prescriptions we used the amplitude of libration can increase
and the system can escape from resonance.
The main difference between both prescriptions comes from the fact that with
\citet{lee_dynamics_2002} prescriptions, the eccentricity damping does not affect
the semi-major axes, while in our model the eccentricity damping terms
contribute to the semi-major axes evolution (see Eq.~(\ref{eq:diskTa})).
Disk-planet interaction models suggest that the semi-major axes evolution are
indeed influenced by the eccentricity damping effect of the disk
\citep[see][]{goldreich_overstable_2014}.
We thus follow these prescriptions in our study.

\subsection{HD~45364b, c: 3:2 MMR, large amplitude of libration}
\label{sec:hd45364}

\tabIII

\figIV

The star \object{HD~45364} hosts two planets
\citep{correia_harps_2009} embedded
in a 3:2 MMR.
The forced eccentricity ratio is
\begin{equation}
 \left(\frac{e_1}{e_2}\right)_{ell} \approx 2.5.
\end{equation}
A large amplitude of libration is observed
\citep[$\sim 70\degr$, see][]{correia_harps_2009},
thus, the stability constraint
gives (Eq.~(\ref{eq:constBigAbis}))
\begin{equation}
  \tau \gtrsim  \left(\frac{e_1}{e_2}\right)_{ell}^2 \approx 6.3.
\end{equation}
The condition of convergent
migration is fulfilled (see Eq.~(\ref{eq:constMigConv})).
The surface density profile exponent is  (see Eq.~(\ref{eq:deftau}))
\begin{equation}
  \beta_\Sigma \lesssim -1.6.
\end{equation}
We obtain again a negative value that corresponds to an inverted profile.
We have
(see Eqs.~(\ref{eq:deltaestim}), (\ref{eq:C1}), and (\ref{eq:C2}))
\begin{eqnarray}
  \delta &=& 6.0\times 10^{-3},\\
  C_1 &=& 0.93,\\
  C_2 &=& 2.3.
\end{eqnarray}
Using these values and $\tau \gtrsim 6.3$, we obtain
(see Eq.~(\ref{eq:constK}))
\begin{equation}
  K \gtrsim 9.4,
\end{equation}
which corresponds to an aspect ratio of (see Eq.~(\ref{eq:defK}))
\begin{equation}
  \frac{H}{a} \lesssim 0.33.
\end{equation}

We performed numerical simulations
with different values of $\tau$
and $K$ (given by Eq.~(\ref{eq:constK})).
The semi-major axes are initially 10 and 14~AU (period ratio of about 1.65),
the eccentricities are 0.001 with anti-aligned periastrons and
coplanar orbits.
The planets are initially at periastrons (zero anomalies).
The evolution of the semi-major axes, the period ratio, the eccentricities,
the eccentricity ratio, and the angles are shown in Fig.~\ref{fig:IV}.
According to our simulations, the amplitude of libration increases
for $\tau \lesssim 10$ (transition between 5-15, see Fig.~\ref{fig:IV})
which is comparable with our analytical result ($\tau\lesssim 6.3$).

It may seem surprising that the amplitude of libration does not increase much more rapidly
for $\tau=2$ than for $\tau=5$ (see Fig.~\ref{fig:IV}).
Indeed, our study shows that the smaller $\tau$ is,
the more unstable the resonant configuration is (Eq.~(\ref{eq:diskAtot})).
However, the evolution of the amplitude of libration does not only depend on the
ratio $\tau = T_{e,1}/T_{e,2}$ but also on the absolute values of
these damping timescales (see Eq.~(\ref{eq:diskAtot})).
In our simulations, we fixed the migration timescale for the outer planet ($T_{m,2}$)
and varied the three other timescales: $T_{m,1} = \tau T_{m,2}$,
$T_{e,2} = T_{m,2}/K$,
$T_{e,1} = \tau T_{m,2}/K$.
The damping timescales are thus much longer for the simulation with
$\tau=2$ ($K=3.5$) than for $\tau=5$ ($K=8$), in order to reproduce the
same equilibrium eccentricities.
This tends to slow down the increasing of the amplitude of libration
and compensates the acceleration provided by decreasing $\tau$.

\section{Conclusion}
\label{sec:conclusion}

We obtained a simple analytical criterion
for the stability of the resonant configuration between two planets
during their migration in a protoplanetary disk.
We used the simplified integrable model of mean-motion resonances
that we developed in
\citet{delisle_resonance_2014},
and modeled the dissipative effect of the disk on the planets by four distinct
timescales: $T_{m,1}$, $T_{m,2}$ (migration of both planets),
and $T_{e,1}$, $T_{e,2}$ (damping of both eccentricities).
As shown by \citet{lee_dynamics_2002}, migrating planets that are captured in
resonance have their eccentricities excited by the migration forces of the disk.
The eccentricities reach equilibrium values between the migration and damping forces.
However, this equilibrium can be unstable, in which case the amplitude of libration
in the resonance increases until the system crosses the separatrix
and escapes from resonance \citep{goldreich_overstable_2014}.
We showed here that the equilibrium is stable on the condition that
$T_{e,1}/T_{e,2} > (e_1/e_2)^2_{ell}$ (ratio of equilibrium eccentricities).
For observed resonant systems, it is probable that the equilibrium was stable
during the migration phase. Otherwise, the planets would have escaped from resonance.
This result allows to put constraints on the damping forces undergone by the planets.
For instance, using prescriptions for type I migration, we show that a locally
inverted profile is needed for resonant systems for which the inner planet
angular momentum deficit (AMD) is larger than the outer planet AMD.
We applied our analytical criterion to \object{HD~60532}b, c (3:1 MMR),
\object{GJ~876}b, c (2:1 MMR), and \object{HD~45364}b, c (3:2 MMR).
We showed for all studied systems
that if the planets had undergone type I migration,
an inverted density profile would be required
for the resonant configuration to be stable.
All these planets are Jupiter-mass planets and are thus believed to open a gap
in the disk and undergo type II migration.
Our results confirm that classical type I migration cannot reproduce the observed systems.

Our model is very general and is not restricted to type I migration.
Considering a scenario of type II migration which is much more realistic for the studied systems,
our model still gives constraints on the migration process and especially
on the eccentricity damping undergone by each planet.
However, we could not find a simple analytical prescription for type II migration
in the literature and thus could not derive constraints on the disk properties in this case.
Having analytical prescriptions for type II migration would allow a more detailed analysis of these
systems.

It would also be very interesting in the future to study small planets in resonance
(with precise enough determination of orbital parameters to be sure of the resonant motion).
Indeed, for small planets, a type I migration scenario is more realistic.
In this case, a local inversion of the density profile (as needed for the three systems
of this study) would be more surprising.

\begin{acknowledgements}
  We thank Rosemary Mardling, Yann Alibert, Wilhelm Kley,
  and the anonymous referee for useful advice.
  This work has been in part carried out within the frame of
  the National Centre for Competence in Research PlanetS
  supported by the Swiss National Science Foundation.
  This work has been supported by PNP-CNRS, CS of Paris
  Observatory, and PICS05998 France-Portugal program.
  JBD acknowledges the financial support of the SNSF.
  AC acknowledges support from CIDMA strategic project UID/MAT/04106/2013.
\end{acknowledgements}

\bibliographystyle{aa}
\bibliography{DCL}

\appendix

\section{Renormalization of coordinates}
\label{sec:renorm}

The renormalized variables are constructed by dividing all actions by the
following constant of motion  \citep[see][]{delisle_dissipation_2012,delisle_resonance_2014}
\begin{equation}
  \label{eq:Gamma}
  \Gamma = \frac{k_2}{k_1}\hat{\Lambda}_1 + \hat{\Lambda}_2,
\end{equation}
where
$\hat{\Lambda}_i = \beta_i\sqrt{\mu_i a_i}$ is the circular angular momentum of the planet $i$.
Noting with a hat the initial actions, the renormalized ones are defined by
\begin{eqnarray}
  \Lambda_i &=& \frac{\hat{\Lambda}_i}{\Gamma},\\
  G_i &=& \frac{\hat{G}_i}{\Gamma},\\
  I_i &=& \frac{\hat{I}_i}{\Gamma},\\
  \mathcal{D} &=& \frac{\hat{\mathcal{D}}}{\Gamma},\\
  \delta &=& \frac{\hat{\delta}}{\Gamma}.
\end{eqnarray}
Expressions~(\ref{eq:La1alpha}) and (\ref{eq:La2alpha}) are straightforwardly derived from these definitions.

Note that the Hamiltonian and the time also have to be renormalized \citep[see][]{delisle_dissipation_2012,delisle_resonance_2014}
in order to preserve Hamiltonian properties.
However, in this study, we consider dissipative forces that act on the system on
very long timescales.
As long as the conservative timescale remains short compared to the dissipation timescale,
the long term evolution of the system is well described by the mean effect
of the dissipation over the conservative timescale.
Therefore, the rescaling of this conservative timescale will not influence the long term evolution of the system.

\section{Reducing to an integrable problem}
\label{sec:simplif}

In the general case, the motion of two planets in a $k_2$:$k_1$ resonance
is described by two degrees of freedom, i.e. both resonant angles
\begin{eqnarray}
  \sigma_1 &=& \frac{k_2}{q} \lambda_2 - \frac{k_1}{q} \lambda_1 - \varpi_1,\\
  \sigma_2 &=& \frac{k_2}{q} \lambda_2 - \frac{k_1}{q} \lambda_1 - \varpi_2,
\end{eqnarray}
and both actions $I_1$, $I_2$.
Let us note $x_i$ the complex cartesian coordinates associated to these action-angle coordinates
\begin{equation}
  x_i = \sqrt{I_i} \expo{\ImUnit \sigma_i}.
\end{equation}
The simplifying assumption introduced in \citet{delisle_resonance_2014} allows to reduce this
generally non-integrable problem to a single degree of freedom one (thus integrable).
The only remaining resonant angle is $\theta$ and the associated action is $\mathcal{D}$.
Noting $u$ the related complex cartesian coordinate
\begin{equation}
  \label{eq:uDtheta}
  u = \sqrt{\mathcal{D}}\expo{\ImUnit \theta},
\end{equation}
the simplifying assumption reads
\begin{eqnarray}
  \label{eq:assumu}
  0 &=& \cos\phi \expo{-\ImUnit\sigma_{2,ell}} x_2 - \sin\phi \expo{-\ImUnit\sigma_{1,ell}} x_1,\\
  \label{eq:u}
  u &=& \cos\phi \expo{-\ImUnit\sigma_{1,ell}} x_1 + \sin\phi \expo{-\ImUnit\sigma_{2,ell}} x_2,
\end{eqnarray}
where $\phi$, $\sigma_{1,ell}$, $\sigma_{2,ell}$ are constant angles defined such that the libration
center is directed toward $u$ \citep[see][]{delisle_resonance_2014}.
Equation~(\ref{eq:u}) shows how the simplified one degree of freedom model mixes both initial degrees of freedom,
and especially, how the resonant angle $\theta$ mixes both initial resonant angles $\sigma_1$, $\sigma_2$.
Equation~(\ref{eq:assum}) directly results from Eq.~(\ref{eq:assumu}).

\section{Evolution of the parameter $\delta$ under dissipation}
\label{sec:evo-delta}

In this section we show how to compute the evolution of the parameter $\delta$ (Eq.~(\ref{eq:ddeltadt})) under
a dissipation affecting the semi-major axes and eccentricities of the planets.
The evolution of the renormalized circular angular momenta only depends on $(\alpha/\dot{\alpha})_d$.
These renormalized quantities are constructed such that
\citep[see Appendix~\ref{sec:renorm} and][]{delisle_resonance_2014}
\begin{equation}
  \frac{\Lambda_1}{\Lambda_2} = \frac{\beta_1\sqrt{\mu_1}}{\beta_2\sqrt{\mu_2}}\sqrt{\alpha}
  \approx \frac{m_1}{m_2}\sqrt{\alpha},
\end{equation}
and
\begin{equation}
  \frac{k_2}{k_1}\Lambda_1 + \Lambda_2 = 1.
\end{equation}
We deduce
\begin{eqnarray}
  \label{eq:dLadt}
  \dot{\Lambda}_1|_d &=& \frac{\Lambda_1\Lambda_2}{2}  \left.\frac{\dot{\alpha}}{\alpha}\right|_d,\\
  \dot{\Lambda}_2|_d &=& - \frac{k_2}{k_1} \frac{\Lambda_1\Lambda_2}{2} \left.\frac{\dot{\alpha}}{\alpha}\right|_d.
\end{eqnarray}
The evolution of $\epsilon$ is then straightforward (see Eq.~(\ref{eq:epsilon}))
\begin{equation}
  \label{eq:depsdt}
  \dot{\epsilon}|_d  = \dot{\Lambda}_1|_d + \dot{\Lambda}_2|_d
  = - \frac{q}{k_1} \frac{\Lambda_1\Lambda_2}{2} \left.\frac{\dot{\alpha}}{\alpha}\right|_d.
\end{equation}
The evolution of the renormalized deficit of angular momentum $I_i$
is given by
\begin{equation}
  \label{eq:dDidt}
  \left.\frac{\dot{I}_i}{I_i}\right|_d = 2 \left.\frac{\dot{\xi}_i}{\xi_i}\right|_d
  + \left.\frac{\dot{\Lambda}_i}{\Lambda_i}\right|_d,
\end{equation}
\begin{equation}
  \label{eq:dD1dt}
  \left.\frac{\dot{I}_1}{I_1}\right|_d = 2 \left.\frac{\dot{\xi}_1}{\xi_1}\right|_d
  + \frac{\Lambda_2}{2} \left.\frac{\dot{\alpha}}{\alpha}\right|_d,
\end{equation}
\begin{equation}
  \label{eq:dD2dt}
  \left.\frac{\dot{I}_2}{I_2}\right|_d = 2 \left.\frac{\dot{\xi}_2}{\xi_2}\right|_d
  - \frac{k_2}{k_1} \frac{\Lambda_1}{2} \left.\frac{\dot{\alpha}}{\alpha}\right|_d
  = 2 \left.\frac{\dot{\xi}_2}{\xi_2}\right|_d
  + \frac{\Lambda_2-1}{2} \left.\frac{\dot{\alpha}}{\alpha}\right|_d.
\end{equation}
We thus deduce
\begin{eqnarray}
  \label{eq:dDdt}
  \left.\frac{\dot{\mathcal{D}}}{\mathcal{D}}\right|_d &=&
  \cos^2\phi \left.\frac{\dot{I}_1}{I_1}\right|_d +
  \sin^2\phi \left.\frac{\dot{I}_2}{I_2}\right|_d\nonumber\\
  &=& 2 \left(\cos^2\phi \left.\frac{\dot{\xi}_1}{\xi_1}\right|_d
    + \sin^2\phi \left.\frac{\dot{\xi}_2}{\xi_2}\right|_d \right)\\
  &&+ \frac{\Lambda_2 - \sin^2\phi}{2} \left.\frac{\dot{\alpha}}{\alpha}\right|_d.\nonumber
\end{eqnarray}
Finally, we have
\begin{eqnarray}
  \dot{\delta}|_d &=& \dot{\mathcal{D}}|_d - \dot{\epsilon}|_d\nonumber\\
  &=& 2 \left(\cos^2\phi \left.\frac{\dot{\xi}_1}{\xi_1}\right|_d
    + \sin^2\phi \left.\frac{\dot{\xi}_2}{\xi_2}\right|_d \right) \mathcal{D}\nonumber \\
  && + \frac{\Lambda_2 - \sin^2\phi}{2} \left.\frac{\dot{\alpha}}{\alpha}\right|_d \mathcal{D}\\
  && + \frac{q}{k_1} \frac{\Lambda_1\Lambda_2}{2} \left.\frac{\dot{\alpha}}{\alpha}\right|_d.\nonumber
\end{eqnarray}

\end{document}